\documentstyle[editedbook,epsfig,psfig,epsf]{mq}

\def\msunyr{M$_{\odot}$ yr$^{-1}$ }

\def\magdot{\ifmmode {}^{\rm m}\!\!\!.\, \else ${}^{\rm m}\!\!\!.\,$\fi}
\def\daydot{\ifmmode {}^{\rm d}\!\!\!.\, \else ${}^{\rm d}\!\!\!.\,$\fi}

\def\cm2{cm$^2$ }
\def\se1{s$^{-1}$ }

\begin{opening}
\title{The central region in SS\,433 supercritical disk and
 origin of flares}
\author{S. Fabrika$^1$, \& T. Irsmambetova$^{2}$ }
\institute{$^1$ Special Astrophysical Observatory, Nizhnij Arkhyz,
      369167, Russia.\\
$^2$ Crimean Lab. Sternberg Astronomical Institute, Nauchny,
       98409, Ukraine.}
\end{opening}

\runningtitle{The central region in SS\,433}
\runningauthor{Fabrika \& Irsmambetova}

\begin{document}
\vspace{-0.5cm}
\begin{abstract}
{\small
Mean orbital light curves of SS433 in different precessional
phases are analysed for active and passive states separately.
In passive states the mean
brightness depends strongly on the disk orientation, the star
is fainter by a factor $\approx 2.2$ in the disk edge-on positions.
In active states the brightness does not depend significantly on
the precessional phase. We suggest that in active states hot gas
cocoons surrounding the inner jets grow and
can not be shielded by the disk rim in the edge-on phases.
Brightest optical flares are clear separated in two groups in orbital
phases, it is considered as indication of orbital eccentricity.
Bright flares prefer specific precession and nodding phases, it
favours the slaved disk model and the flares as disk perturbations by
a torque applied to outer parts of the accretion disk.}
\end{abstract}

\section{Active and passive states of SS\,433}

Active states of SS\,433 were isolated using the GBI radio monitoring
program data (http://www.gb.nrao.edu/fgdoss/gbi/gbint.html) and direct
inspection
\vspace*{-0.5cm}
\begin{figure}[h]
\centering \epsfig{file=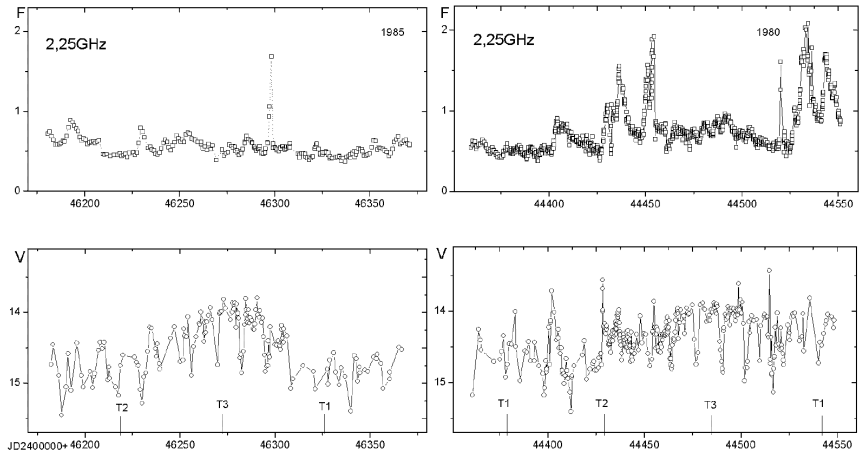,width=11.5cm}
\vspace*{-0.9cm}
\caption{
Radio and optical data in 1985 when SS\,433 was mainly in
quiet state and in 1980 when it was mainly active. The radio flux
(GBI radio monitoring data) in Jy, optical flux in V magnitudes}
\label{fig1}
\end{figure}
of the optical data.
Fig.\,1 shows radio and optical data in two
observational seasons
(1980 and 1985). Active states are clearly seen in radio.
In visible region flares destroy the regular orbital and precessional
variabilities. Orbital variability is seen
\begin{figure}[htb]
\vspace*{-0.4cm}
\centering \epsfig{file=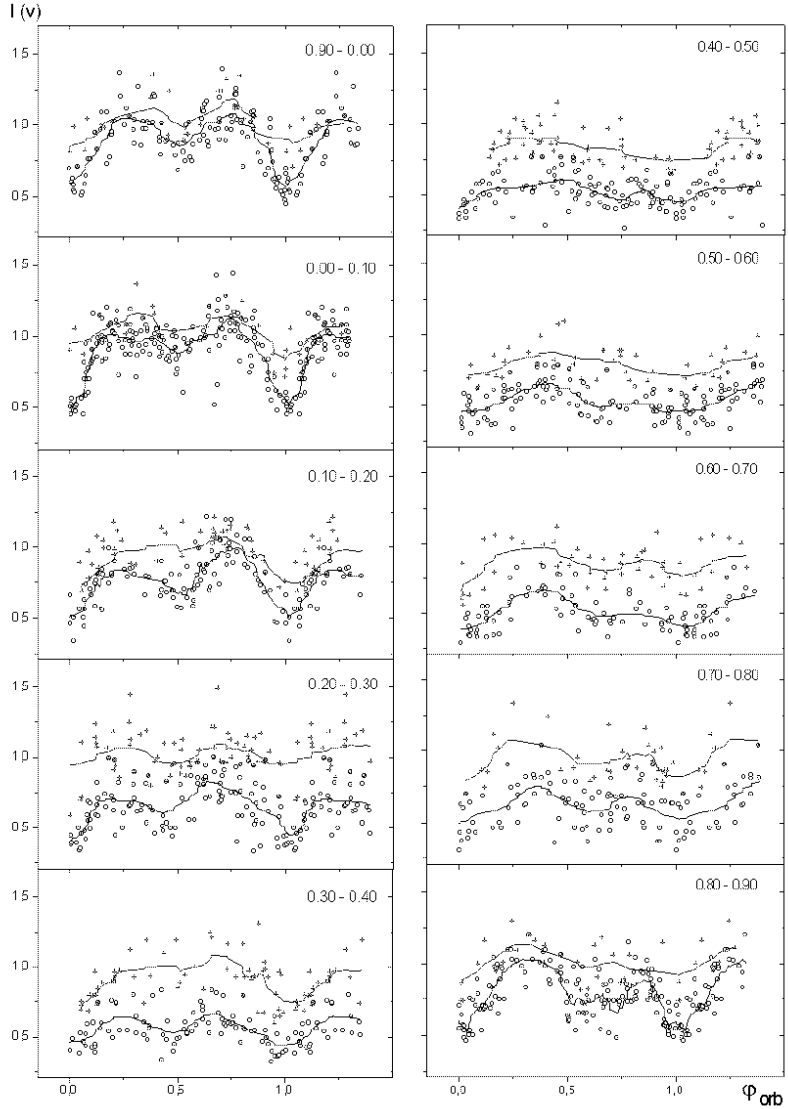,width=11cm}
\caption{
Mean orbital light curves for passive (circles, down curves)
and active (crosses, upper curves)
states in different phases of precession. Obvious flares were excluded.
Relative intensity I\,$=0$ corresponds to V\,=\,14\magdot0. The accretion
disk eclipses are phased at $\phi_{orb}=0$}
\label{fig2}
\end{figure}
as deep primary eclipses of the accretion
disk by the donor--star (Min\,I, $\phi_{orb}=0$, $P_{orb}=13\daydot08$).
Precessional variability is a brightening when the accretion
disk is the most open to observer ($T_3$ moment, $\psi_{pr}=0$,
$P_{pr}=162\daydot4$) and weakening when the disk is in edge--on
positions ($T_{1,2}$ moments, $\psi_{pr}=0.34, 0.66$).

We analyse the orbital light curves of SS433 in different precessional
phases for active and passive states separately.
Both original and all published data of optical V--band
photometry for 1979--1996 were used. The data--base consists of
2200 individual observations collected in Sternberg Institute.
We used 1491 observations in passive states of SS\,433 and 584
observations in active states, where obvious flares were excluded.
We find that the light curve in active state is about the same as
that of in passive with primary and secondary minima (Fig.\,2, Fig.\,3).
However, it is very important that in active states the brightness
does not depend significantly on precessional phase and the primary
minima are not so deep
as they are in passive states. We suggest a geometry of the inner
disk parts as
two hot gas cocoons surrounding the two inner jets. In active states the
cocoons grow and they can not be shielded by the disk rim when the disk
is edge--on. In Fig.\,4 we show a sketch of the disk and cocoons
in active and passive periods.

\begin{figure}[htb]
\vspace*{-0.5cm}
\centering \epsfig{file=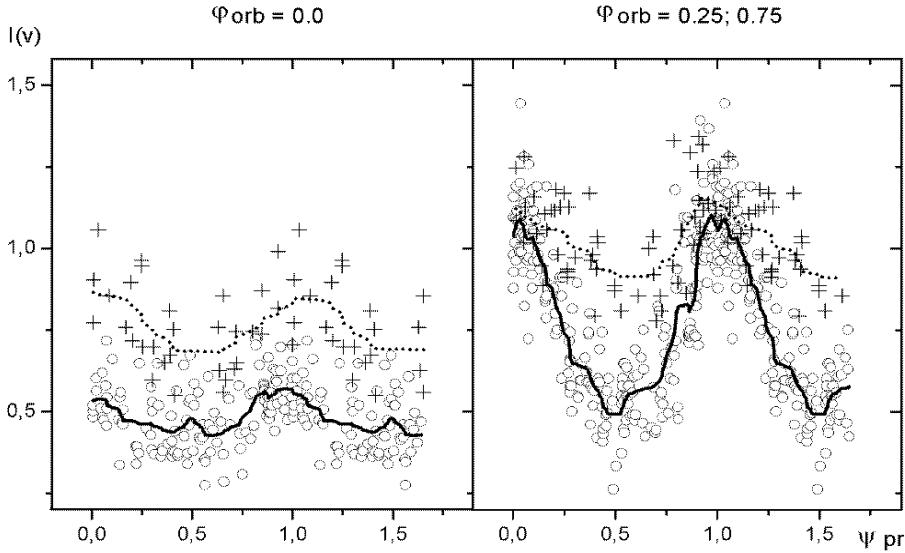,height=8cm,width=12.5cm}
\caption{
Mean precessional light curves for passive (circles, down curves) and
active (crosses, upper curves) states in a middle
of the primary minimum (left) and in elongations (right) in phase intervals
$\Delta \phi = \pm 0.05$}
\label{fig3}
\end{figure}

When the disk is the most open to observer the mean brightness in
elongations is the same in active and passive states. Probably the cocoon
surrounding the approaching jet
is not shaded up to its base by the disk rim in these precession phases
and luminosity of the cocoon does not depend notably on its size.
This may be in a case if the cocoon scatters ($\tau_T \sim 1$)
inner radiation coming from
the accretion disk funnel. The cocoons can be identified with a source of
the UV radiation of SS\,433, where Dolan et al. \cite{Dea97}
have detected
the strong linear polarization directed along jets. The cocoons can be
also identified with a source of the double--peaked He\,II\,$\lambda 4686$
line observed in the disk \cite{F97}.

\begin{figure}[htb]
\centering \epsfig{file=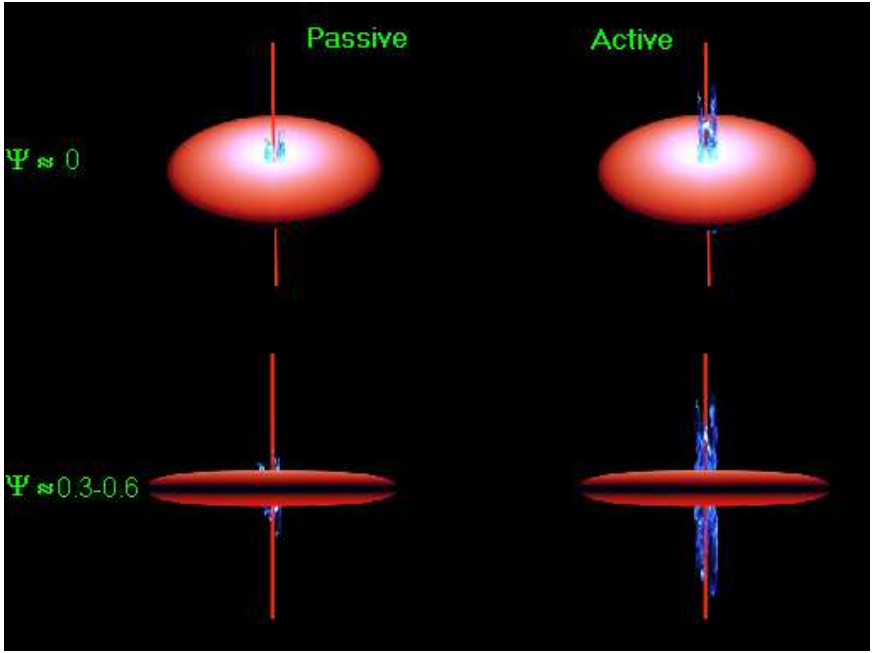,width=11.5cm}
\caption{
A sketch of the accretion disk with cocoons surrounding inner jet bases
in active and passive states in two extreme precessional orientations}
\label{fig4}
\end{figure}

In passive states an amplitude of precessional modulation
($\Delta I \approx 0.4$) is about the same as amplitute of primary
minima ($\Delta I \approx 0.5$). This means that the projected sizes of
the outer disk rim and the companion star are the same.

\section{Phasing of flares}

We have selected optical flares as short (about one day) increases
of brightness over the mean passive state flux in corresponding phases.
14 super--bright flares ($\Delta I_{f}= 0.7-1.2$) and 16 bright flares
($\Delta I_{f}= 0.57-0.7$) were isolated and studied in the
phase diagrams (Fig.\,5).
Fainter flares show about the same behaviour as bright ones,
but their distributions are more scattered.

All the super flares (save for one) are located in two isolated
orbital phases.  The only possibility to produce flares in
specific orbital phase is a noncircular orbit. We suggest that it
is the case. If a periastron passage occurs close to $\phi_{orb}
\approx 0.28$, the second group of flares delays for
$\Delta \phi \approx 0.1$ to expected apoastron passage ($\phi_{orb}
\approx 0.78$). The mass transfer rate in SS\,433 is highly
supercritical $\dot M \sim
10^{-4}$\,{\msunyr} \cite{vdH81}, and flares have to
be related not to the mass transfer rate variations, but rather
to the accretion disk perturbations.

\begin{figure}[htb]
\vspace*{-0.5cm}
\centering \hspace*{-0.0cm} \epsfig{file=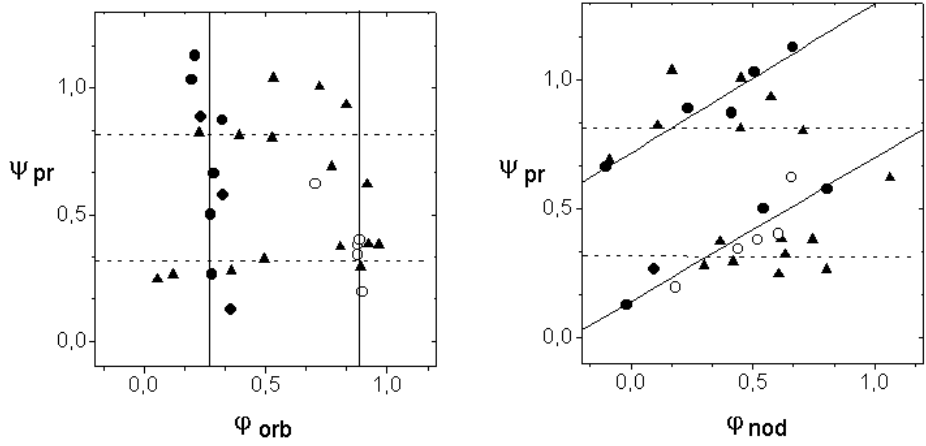,width=14cm}
\vspace*{-1.7cm}
\caption{
Phase diagrams for super--bright flares (filled and open circles)
and bright flares (triangles). Two vertical lines mark location
of the super--flares
at $\phi_{orb} = 0.28 \pm 0.05$ and $0.886 \pm 0.004$, two horizontal
lines mark location of the cluster of bright flares at $\psi_{pr} =
0.32 \pm 0.05$ and $+0.5$ in precession phase. In the right diagram
with nodding phases ($\omega_{nod} = 2\omega_{orb} + 2\omega_{pr}$)
two diagonal lines represent the best linear fits for super flares
only with slopes 0.57($\pm9$) and 0.58($\pm9$)}
\label{fig5}
\end{figure}

The eccentricity in SS\,433 is less than $e < 0.05$ \cite{FKS90}.
However even an eccentricity $e \approx 0.01$ can produce the critical
Roche volume variation of 2\,$\%$ \cite{BG84}. Such an
eccentricity may dominate the effect of the primary's Roche volume
variation appearing because of primary's spin misalignment with the
orbital axis (twice per orbit, the slaved disk model).

A real moment of the periastron passage has to be earlier than the
flares on
the time of matter transfer through the disk ($\Delta t_{tr}$),
$\phi_{orb} = 0.28 - \Delta \phi_{tr}$. The specific orbital phase
specifies precession phases for flares depending on mechanism of
flares. They are $\psi_{pr} \approx 0.22(0.72) + \Delta \phi_{tr}$,
if perturbations occur because of (i) a torque applied to outer parts
of accretion disk (the nodding motions, \cite{KGM82}) or
$\psi_{pr} = 0.47(0.97) + \Delta \phi_{tr}$, if the perturbations occur
because of (ii) the Roche lobe squeezing \cite{BG84}.
Observed location of the super flares
allows for any of the two models, however the bright flares and some super
flares are crowded in precessional phase $\psi_{pr} \approx 0.3$, what
favours the nodding model (i), if $\Delta t_{tr} \sim 1$ day.

In the diagram $\psi_{pr} - \phi_{nod}$ (Fig.\,5) the super
flares show
quite expected behaviour as flares in fixed orbital phases ($\psi_{pr}
\propto 0.5 \, \phi_{nod}$). The phases $\phi_{nod}=0.25,0.75$ correspond here
to location of companions in the line of nodes. The bright flares,
which do not
follow the diagonal lines, delay for $1.0 \pm 0.4$ days from nodding phases
$\phi_{nod} = 0.0(0.5)$. Such a behaviour could be understood, if the
nodding (i) mechanism does work and the time of matter transfer
across the perturbed disk is $\Delta t_{tr} \sim 1$ day.

\section*{Acknowledgments}
The authors thank G.\,Valyavin for a help in preparation of the paper.
This work has been partly supported by the RFBR grant N\,00-02-16588
and the Russian Federal Program ``Astronomy''.

\end{document}